\documentclass[sigconf]{acmart}

\usepackage{amsmath,amsfonts}
\usepackage{algorithmic}
\usepackage{graphicx}
\usepackage[ruled, vlined]{algorithm2e}
\usepackage{textcomp}
\usepackage{multirow}
\usepackage{hyperref}
\usepackage{caption}
\usepackage{outlines}
\usepackage{soul}
\usepackage{booktabs}  
\usepackage{subcaption}

\newtheorem{claim}{Claim}

\newtheorem*{claim1}{Claim 1 restated}
\newtheorem*{claim2}{Claim 2 restated}

\copyrightyear{2022} 
\acmYear{2022} 
\setcopyright{rightsretained} 
\acmConference[WPES '22]{Proceedings of the 21st Workshop on Privacy in the Electronic Society}{November 7, 2022}{Los Angeles, CA, USA}
\acmBooktitle{Proceedings of the 21st Workshop on Privacy in the Electronic Society (WPES '22), November 7, 2022, Los Angeles, CA, USA}
\acmDOI{10.1145/3559613.3563198}
\acmISBN{978-1-4503-9873-2/22/11}


\begin{document}
\title{SplitGuard: Detecting and Mitigating Training-Hijacking Attacks in Split Learning}

\author{Ege Erdoğan}
\affiliation{%
  \institution{Koç University}
  \city{\.Istanbul}
  \country{Turkey}
}
\email{eerdogan17@ku.edu.tr}

\author{Alptekin Küpçü}
\affiliation{%
  \institution{Koç University}
  \city{\.Istanbul}
  \country{Turkey}
}
\email{akupcu@ku.edu.tr}

\author{A. Ercüment Çiçek}
\affiliation{%
  \institution{Bilkent University}
  \city{Ankara}
  \country{Turkey}
}
\email{cicek@cs.bilkent.edu.tr}

\begin{abstract}
Distributed deep learning frameworks such as \textit{split learning} provide great benefits with regards to the computational cost of training deep neural networks and the privacy-aware utilization of the collective data of a group of data-holders. Split learning, in particular, achieves this goal by dividing a neural network between a client and a server so that the client computes the initial set of layers, and the server computes the rest. However, this method introduces a unique attack vector for a malicious server attempting to steal the client's private data: the server can direct the client model towards learning any task of its choice, e.g. towards outputting easily invertible values. With a concrete example already proposed (Pasquini et al., CCS '21), such \textit{training-hijacking} attacks present a significant risk for the data privacy of split learning clients. 

In this paper, we propose SplitGuard, a method by which a split learning client can detect whether it is being targeted by a training-hijacking attack or not. We experimentally evaluate our method's effectiveness, compare it with potential alternatives, and discuss in detail various points related to its use. We conclude that SplitGuard can effectively detect training-hijacking attacks while minimizing the amount of information recovered by the adversaries.
\end{abstract}

\maketitle

\section{Introduction}

Training deep neural networks (DNNs) requires large amounts of computing power and data; however, relying on a sustained increase in computing power is unsustainable \cite{thompson2020computational}, and data from multiple sources cannot always be aggregated (e.g. healthcare data regulations \cite{annas_hipaa_2003, mercuri_hipaa-potamus_2004}).

Distributed deep learning frameworks such as \textit{split learning} (SplitNN) \cite{vepakomma_split_2018, gupta_distributed_2018} and \textit{federated learning} \cite{bonawitz_towards_2019, konecny_federated_2016, konecny_federated_2017} aim to solve these two problems by allowing a group of data-holders (clients) to train a DNN without raw data sharing. The resulting DNN is effectively trained using the data-holders' collective data. 

In federated learning, each client trains a local model and sends its parameter updates to the central server. The server aggregates the parameter updates (e.g. taking their average) of the clients and redistributes the final value. In SplitNN, a DNN is split into multiple parts (typically two); in the two-part setting, clients compute the first few layers of a DNN and send the output to a central server, who then computes the rest of the layers and initiates the backpropagation of gradients. In both methods, no client shares its private data with another party, and all clients end up with the same model.

\textbf{The Problem.} In SplitNN, the server has control over what the client models learn since it propagates the parameter updates back to the clients. This creates a new attack vector we call \textit{training-hijacking}, that has already been exploited in an attack (Pasquini et al., CCS '21) \cite{pasquini_unleashing_2021}, for a malicious server trying to obtain the clients' private data.\footnote{By contrast, this attack vector does not exist in federated learning, since the clients can trivially check if their model is aligned with their goals by calculating its accuracy. Running the same detection method is not possible in split learning since the server can train a legitimate model on the side using the clients' intermediate outputs (i.e. follow the protocol) and use that model for an accuracy test.} In the attack, the server discards the original classification task and leads a client towards outputting values in such a way that it is as easy as possible for the server to obtain back the original inputs from the intermediate values. This is a serious potential violation of the clients' data privacy, but if the clients can detect early in the training process that the server is launching an attack, they can halt training and leave the attacker empty-handed.

\textbf{Our Solution.} In this paper we propose SplitGuard, a protocol by which a SplitNN client can detect, without expecting cooperation from the server, if its local model is being hijacked. To the best of our knowledge, SplitGuard is the first attempt at detecting training-hijacking attacks. Our starting point is the observation that if a client's local model is learning the intended classification task, then it should behave in a drastically different way when the task is reversed (i.e. when success in the original task implies failure in the new task). In classification, this reversal means trying to learn a training batch with random label values. We demonstrate through various experiments (using the MNIST \cite{lecun2010mnist}, Fashion-MNIST \cite{xiao2017/online}, and CIFAR10/100 \cite{Krizhevsky09learningmultiple} datasets) that the emergence of this discrepancy precedes the server being able to extract useful information, effectively giving the clients the upper hand against the attackers.

The code for our approach is available at \url{https://github.com/ege-erdogan/splitguard}.

\begin{figure*}[t!]
    \centering
    \begin{subfigure}{0.48\textwidth}
        \centering
        \includegraphics[width=\textwidth]{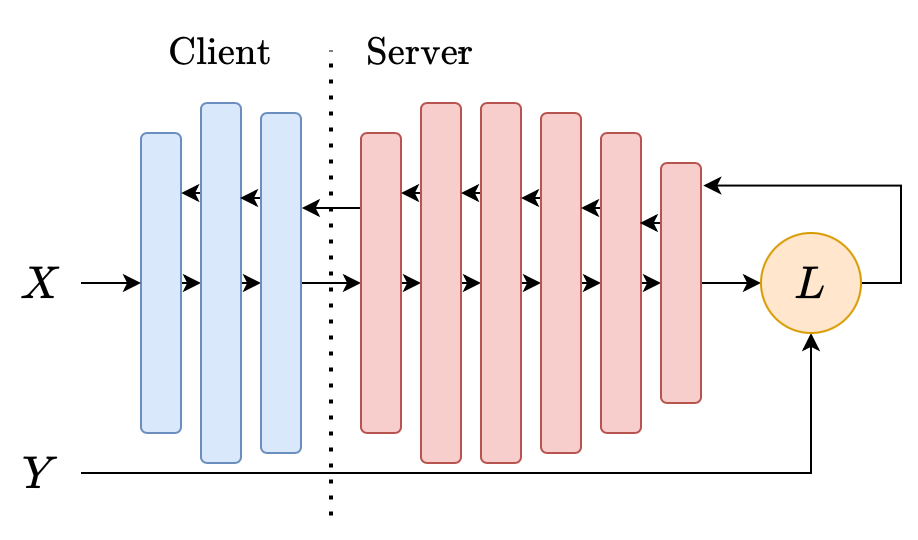}
        \caption{With label-sharing.}
        \label{fig:splitnn_label_sharing}
    \end{subfigure}
    \begin{subfigure}{0.48\textwidth}
        \centering
        \includegraphics[width=\textwidth]{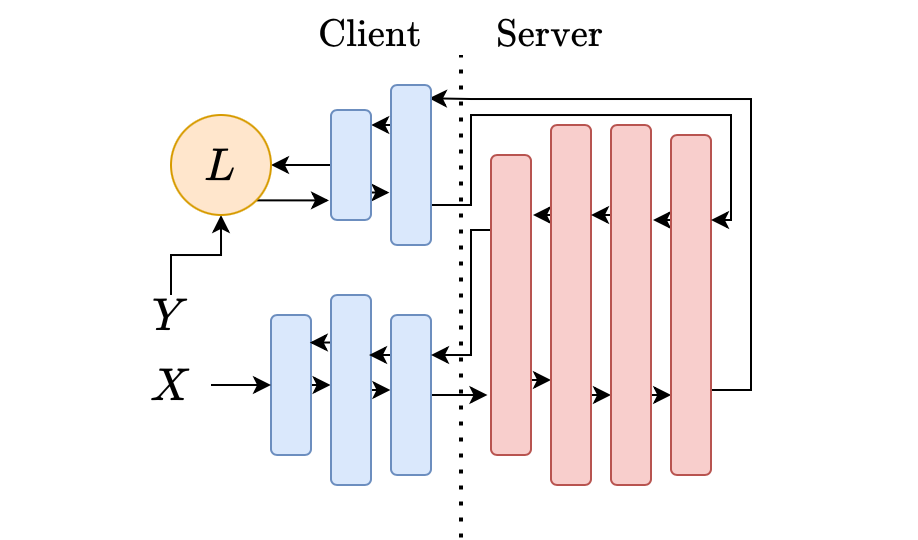}
        \caption{Without label-sharing.}
        \label{fig:splitnn_private_labels}
    \end{subfigure}
    \caption{Different split learning setups. Arrows denote the forward and backward passes, starting with the examples $X$, and propagating backwards after the loss computation using the labels $Y$. In Figure \ref{fig:splitnn_label_sharing}, clients send the labels to the server along with the intermediate outputs. In Figure \ref{fig:splitnn_private_labels}, the model terminates on the client side, and the clients do not share their labels.}
    \label{fig:splitnn_setups}
\end{figure*}

\section{Background and Related Work}

\subsection{Neural Networks}

In the context of supervised learning,\footnote{Supervised learning involves learning through labeled data, as opposed to unsupervised learning, in which the data used is not labeled.} a neural network \cite{Goodfellow-et-al-2016} is a parameterized function $f:X \times \Theta \rightarrow Y$ that approximates a function $f^*:X \rightarrow Y$. The training process aims to learn the parameters $\Theta$ using a training set consisting of examples $\tilde X$ and labels $\tilde Y$ sampled from the real-world distributions $X$ and $Y$.

A typical neural network, also called a \textit{feedforward neural network}, consists of discrete units called \textit{neurons}, organized into layers. Each neuron in a layer takes in a weighted sum of the previous layer's neurons' outputs, applies a non-linear activation function, and outputs the result. The weights connecting the layers to each other constitute the parameters that are updated during training. Considering each layer as a separate function, we can model a neural network as a chain of functions, and represent it as $f(x)=f^{(N)}(...(f^{(2)}(f^{(1)}(x)))$, where $f^{(1)}$ corresponds to the first layer, $f^{(2)}$ to the second layer, and $f^{(N)}$ to the final, or the \textit{output} layer. The final layer often has a different activation function, such as the softmax function.

Training a neural network involves minimizing a loss function. However, since the nonlinearity introduced by the activation functions applied at each neuron causes the loss function to become non-convex, we use iterative, gradient-based approaches to minimize the loss function. Since these methods do not provide any global convergence guarantees, it is important that the training data represent the real-world data as accurately as possible.

A widely-used optimization method is \textit{stochastic gradient descent} (SGD). Rather than computing the gradient from the entire data set, SGD computes gradients for batches selected from the data set. The weights are updated by propagating the error backwards using the backpropagation algorithm. Training a deep neural network generally requires multiple passes over the entire data set, each such pass being called an \textit{epoch}. One round of training a neural network requires two passes through the network: one forward pass to compute the network's output, and one backward pass to update the weights. We will use the terms \textit{forward pass} and \textit{backward pass} to refer to these operations in the following sections. For an overview of gradient-based optimization methods other than SGD, we refer the reader to \cite{ruder_overview_2017}.

\subsection{Split Learning}

In split learning (SplitNN) \cite{gupta_distributed_2018, vepakomma_no_2018, vepakomma_split_2018}, a DNN is split between the clients and a server such that each client locally computes the first few layers, and the server computes rest of the layers. This way, a group of clients can train a DNN utilizing, but not sharing, their collective data. This way, most of the computational work is offloaded to the server, reducing the cost of training for the clients. However, this partitioning involves a privacy/cost trade-off for the clients, with the outputs of earlier layers leaking more information about the inputs. 

Figure \ref{fig:splitnn_setups} displays the two basic setups of SplitNN, the main difference being whether the clients share their labels with the server or not. In Figure \ref{fig:splitnn_label_sharing}, clients compute only the first few layers, and share their labels with the server. The server then computes the loss value, starts backpropagation, and sends the gradients of its first layer back to the client, who then completes the backward pass. The private-label scenario depicted in Figure \ref{fig:splitnn_private_labels} follows the same procedure, with an additional communication step. Since now the client computes the loss value and initiates backpropagation, it should first feed the server model with the gradient values to resume backpropagation. 

The primary advantage of SplitNN compared to federated learning is its lower communication load \cite{singh2019detailed}. While federated learning clients have to share their entire parameter updates with the server, SplitNN clients only share the output of a single layer. 

SplitNN follows a round-robin training protocol to accommodate multiple clients; clients take turn training with the server using their local data. Before a client starts its turn, it should bring its parameters up-to-date with those of the most recently trained client. There are two ways to achieve this: the clients can either share their parameters through a central parameter server, or directly communicate with each other in a P2P way.

Choosing a split depth is crucial for SplitNN to actually provide data privacy. If the initial client model is too shallow, an honest-but-curious server can recover the private inputs with high accuracy, knowing only the model architecture (not the parameters) on the clients' side \cite{erdogan2021unsplit}. This implies that SplitNN clients should increase their computational load by computing more layers for stronger data privacy. 

\subsection{Training-Hijacking in Split Learning}

In a training-hijacking attack against a SplitNN client, the attacker server tries to direct the client models towards its own malicious goal, independent of the actual classification task. The Feature-Space Hijacking Attack (FSHA) (Pasquini et al. CCS '21) \cite{pasquini_unleashing_2021} is the only proposed training-hijacking attack against SplitNN clients so far. The server aims to lead the clients, by propagating back loss values independent of the original task, towards outputting values in such a way that it is easier to recover the original inputs (clients' private data) than if the model was learning the original task.

In FSHA, the attacker (a SplitNN server) first trains an autoencoder (consisting of the encoder $\tilde{f}$ and the decoder $\tilde{f}^{-1}$) on some public dataset $X_{pub}$ similar to that of the client's private dataset $X_{priv}$. It is important for the attack's effectiveness that $X_{pub}$ be similar to $X_{priv}$. Without such a dataset at all, the attack cannot be launched. The main idea is for the server to bring the output spaces of the client model $f$ and the encoder $\tilde f$ as close as possible, so that the decoder $\tilde{f}^{-1}$ can successfully invert the client outputs and recover the private inputs.

After this initial \textit{setup phase}, the client model's training begins. For this step, the attacker initializes a distinguisher model $D$ that tries to distinguish the client's output $f(X_{priv})$ from the encoder's output $\tilde{f}(X_{pub})$. More formally, the distinguisher is updated at each iteration to minimize the loss function.
\begin{equation}
    L_D = \log(1-D(\tilde{f}(X_{pub}))) + \log(D(f(X_{priv}))).
\end{equation}
Simultaneously at each training iteration, the server directs the client model $f$ towards maximizing the distinguisher's error rate, thus minimizing the loss function.
\begin{equation}
    L_f = \log(1-D(f(X_{priv}))).
\end{equation}
In the end, the output spaces of the client model and the server's encoder are expected to overlap to a great extent, making it possible for the decoder to invert the client's outputs. 

Notice that the client's loss function $L_f$ is totally independent of the training labels, as in changing the value of the labels does not affect the loss function. We will soon refer to this observation.

\subsection{Differential Privacy as a Defense Against Training-Hijacking}

Differential privacy \cite{dwork2014algorithmic}, with the intention of minimizing the model's memorization of private training data, can potentially be used as a defense against training-hijacking attacks. In such an attempt, Gawron and Stubbings \cite{gawron2022feature} apply differential privacy on the gradients received from the SplitNN server on the client side, and compare the results a FSHA server obtains with those of the non-DP scenario. Applying DP makes the attack less effective for the same number of iterations, but the attacker still obtains high-accuracy results after a higher number of iterations. 

Thus, differential privacy by itself does not make training-hijacking attacks a nonconcern for SplitNN clients, the bottom line being as the authors state: "DP can at most delay FSHA convergence."

Nevertheless, this delay can still prove useful. For example, a SplitNN client running SplitGuard while also applying DP on the gradients it receives would have more time to detect a training-hijacking attack before the attacker learns something of value. Hence, DP should not be ruled out as a defense against training-hijacking; in fact, it can be a strong tool for clients when used in the right context.

\begin{table}[h!]
    \centering
    \caption{Summary of notation used throughout the paper.}
    \label{tab:notation}
    
    \begin{tabular}{ll}
         \toprule
         \multicolumn{2}{l}{\textbf{Notation}} \\ \midrule
         $P_F$            & Probability of sending a fake batch \\ 
         $B_F$            & Share of randomized labels in a fake batch \\ 
         $N$              & Batch index at which SplitGuard starts running \\ 
         $F$              & Set of fake gradients \\ 
         $R_1, R_2$       & Random, disjoint subsets of regular gradients \\ 
         $R$              & $R_1 \cup R_2$ \\ 
         $\alpha, \beta$  & Parameters of the SplitGuard score function \\ 
         $L$              & Number of classes \\
         $A$              & Model's classification accuracy \\
         $A_F$            & Expected classification accuracy for a fake batch \\
         \bottomrule
    \end{tabular}
\end{table}

\section{SplitGuard}

We start our presentation of SplitGuard by restating an earlier remark: \textit{If the training-hijacking detection protocol requires the attacking SplitNN server to knowingly take part in the protocol, the server can easily circumvent the protocol by training a legitimate model on the side, and using that model during the protocol's run.} In the light of this, it is evident that we need a method which the clients can run during training and without breaking the flow of training from the server's point of view.

\subsection{Overview}

Our main idea is that if the client model is learning the intended task, then it should behave in a drastically different way when that task is reversed (e.g. for classification, when the label values in a training batch are randomly reassigned); since the attacker's objective is independent of the original task, the same discrepancy should not be visible if the server is hijacking the training process. We then need this discrepancy to become evident before the attacker can learn significant information so that the clients can stop training soon enough if the expected discrepancy \textit{does not} occur.

During training with SplitGuard, clients intermittently input batches with randomized labels, denoted \textit{fake batches}, as opposed to \textit{regular batches}.\footnote{\textit{Fake gradients} and \textit{regular gradients} similarly refer to the gradients resulting from fake and regular batches.}   

There are two components of the aforementioned discrepancy between the fake and regular gradients: \textit{angle} and \textit{magnitude}. The client model learning the intended task means it is moving towards a high-accuracy point on its parameter space. When the labels are randomized, that high-accuracy point becomes a low-accuracy point; the model tries to move away from that point, and the classification error increases.\footnote{For an extreme example, suppose that some point in the parameter space corresponds to perfect accuracy. If all the labels in a batch are changed, that point will correspond to $0\%$ accuracy since.} More specifically, we make the following two claims (experimentally validated in Section \ref{validation}):
\begin{claim}
If the client model is learning the intended task, then the angle between fake and regular gradients will be higher than the angle between two random subsets of regular gradients.\footnote{Angle between \textit{sets} meaning the angle between the \textit{sums} of vectors in those sets.}
\end{claim}
\begin{claim}
If the client model is learning the intended task, then fake gradients will have a higher magnitude than regular gradients.
\end{claim}

\subsection{Putting the Claims to Use}

\begin{figure}[t!]
    \centering
    \includegraphics[width=0.5\textwidth]{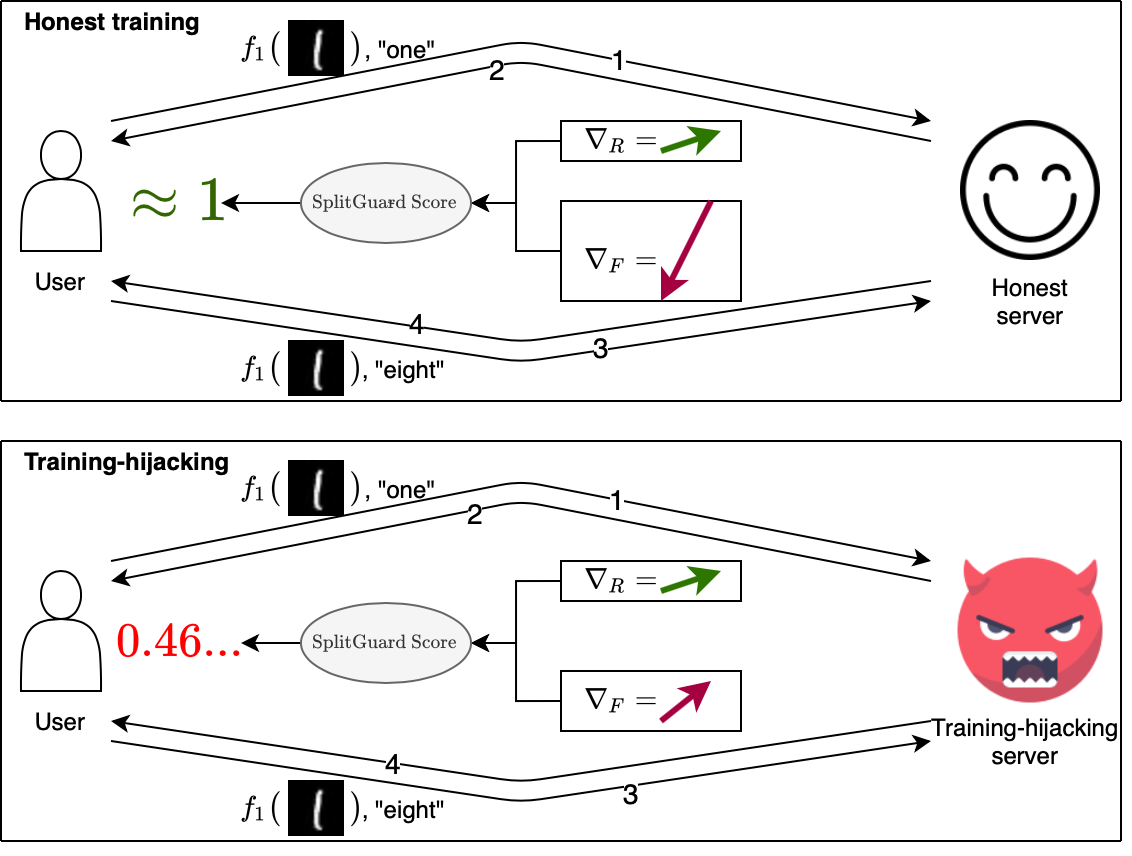}
    \caption{Overview of SplitGuard, comparing the honest training and training-hijacking scenarios. Clients intermittently send training batches with randomized labels (e.g. random value "eight" instead of the true label "one"), and then analyze the behavior of their local models from their parameter updates.}
    \label{fig:splitguard}
\end{figure}

At the core of SplitGuard, clients compute a value, denoted the SplitGuard score, based on the fake and regular gradients they have collected up to that point. This value's history is then used to reach a decision on whether the server is launching an attack or not. We now describe this calculation process in more detail. Table \ref{tab:notation} displays the notation we use from here on.

Starting with the $N$th batch during the first epoch of training, with probability $P_F$,\footnote{This is equivalent to allocating a certain share of the training dataset for this purpose before training.} clients send fake batches in which the share $B_F \in [0,1]$ of the labels are randomized. Upon calculating the gradient values for their first layer, clients append the fake gradients to the list $F$, and split the regular gradients randomly into the lists $R_1$ and $R_2$, where $R=R_1 \cup R_2$. To minimize the effect of fake batches on model performance, clients discard the parameter updates resulting from fake batches. 

Figure \ref{fig:splitguard} displays a simplified overview of the protocol, and Algorithm \ref{alg:main} explains the modified training procedure in more detail. The \texttt{MAKE\_DECISION} function contains the clients' decison-making logic and will be described later in Algorithm \ref{alg:decision}.

\begin{algorithm}[t!]
$f, w$: client model, parameters \\ 
$OPT$: optimizer \\ 
$P_F$: probability of sending fake batches \\ 
$B_F$: number of labels randomized in fake batches \\ 
$N$: number of initial batches to ignore \\
initialize $R_1, R_2, F$ as empty lists \\ 
\texttt{rand}($y$, $B_F$): randomize share $B_F$ of the labels $Y$. \\

\textbf{before} \textit{training}, set parameters $\alpha, \beta, B_F, P_F, T, N$. \\
\While{training}{
    \For{$(x_i,y_i)$ $\leftarrow$ trainset}{
        \uIf{probability $P_{\text{F}}$ occurs \textbf{and} $i \geq N$}{
            // sending fake batches \\
            Send $(f(x_i), \texttt{rand}(y_i, B_F))$ to server \\ 
            Receive gradients $\nabla_F$ from server \\ 
            Append $\nabla_F$ to F \\ 
            \texttt{MAKE\_DECISION}$(F, R_1 \cup R_2)$ \\
            // do not update parameters \\
        }
         \Else {
            // regular training \\
            Send $(f(x_i), y_i)$ to server \\ 
            Receive gradients $\nabla_R$ from server \\ 
            \If{$i \geq N$}{
                \uIf{probability 0.5 occurs}{
                Append $\nabla_R$ to $R_1$
                } \Else {
                    Append $\nabla_R$ to $R_2$
                }
            }
            $w \leftarrow w + OPT(\nabla_R)$ \\
        }
    }
}
\caption{Client training with label sharing}
\label{alg:main}
\end{algorithm}

We should first define two quantities. For two sets of vectors $A$ and $B$, we define $d(A,B)$ as the absolute difference between the average magnitudes of the vectors in $A$ and $B$:
\begin{equation}
    d(A, B) = \Big| \frac{1}{|A|} \sum_{a \in A} \Vert a \Vert -  
                    \frac{1}{|B|} \sum_{b \in B} \Vert b \Vert
                \Big|,
\end{equation}
and $\theta(A,B)$ as the angle between sums of vectors in two sets $A$ and $B$:
\begin{equation}
    \theta(A,B) = arccos\Big(\frac{\bar A \cdot \bar B}{\Vert \bar A \Vert \cdot \Vert \bar B \Vert}\Big)
\end{equation}
where
\begin{equation}
    \bar A = \sum_{a \in A} a
\end{equation}
for a set of vectors $A$. We can restate our two claims more concisely using these quantities under the condition that the client model is learning the intended task:
\begin{claim1}
$\theta(F, R) > \theta(R_1, R_2)$
\end{claim1}
\begin{claim2}
$d(F,R) > d(R_1, R_2)$
\end{claim2}

If the model is learning the intended task, then it follows from the two claims that the product $\theta(F,R) \cdot d(F,R)$ will be greater than the product $\theta(R_1,R_2)\cdot d(R_1,R_2)$. If the model is learning some other task independent of the labels, then $F, R_1$, and $R_2$ will essentially be three random samples of the set of gradients obtained during training, and it will not be possible to consistently detect the same relationships among them.

\begin{figure*}[t!]
    \centering
    \begin{subfigure}{0.24\textwidth}
        \includegraphics[width=\textwidth]{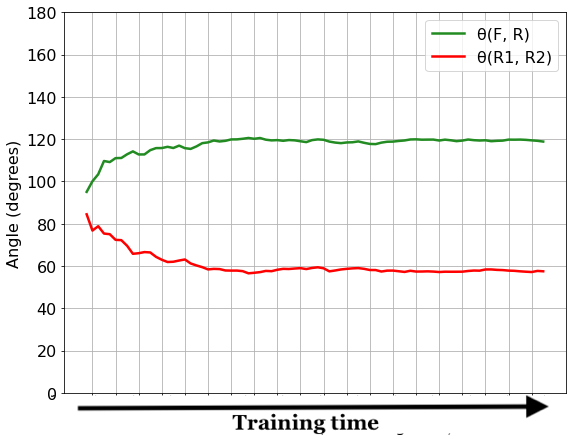}
        \caption{MNIST}
        \label{fig:mnist_angle}
    \end{subfigure}
    \begin{subfigure}{0.24\textwidth}
        \includegraphics[width=\textwidth]{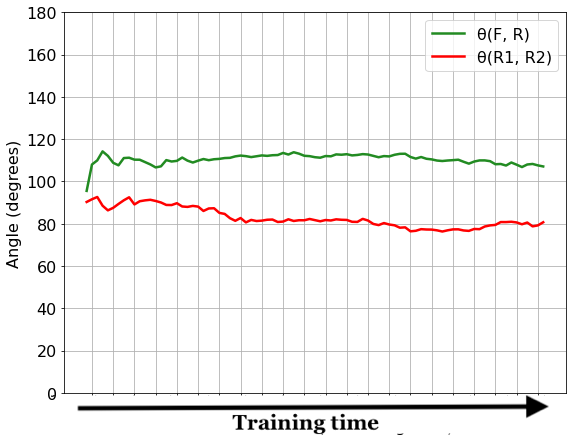}
        \caption{Fashion-MNIST}
        \label{fig:f_mnist_angle}
    \end{subfigure}
    \begin{subfigure}{0.24\textwidth}
        \includegraphics[width=\textwidth]{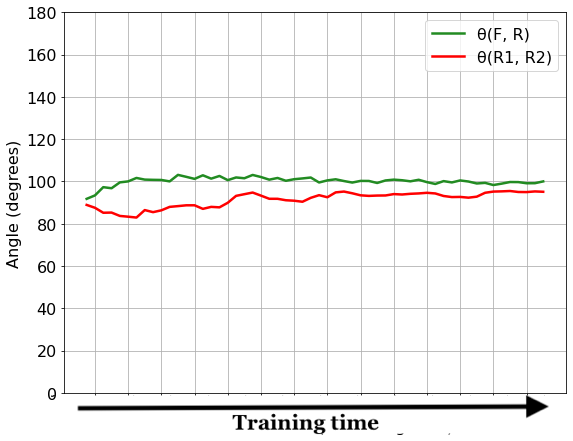}
        \caption{CIFAR10}
        \label{fig:cifar_angle}
    \end{subfigure}
    \begin{subfigure}{0.24\textwidth}
        \includegraphics[width=\textwidth]{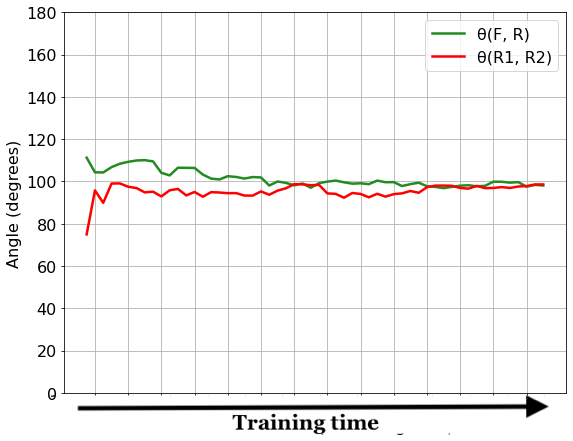}
        \caption{CIFAR100}
        \label{fig:cifar100_angle}
    \end{subfigure}
    \caption{Comparison of the angle between fake and regular gradients ($\theta(F,R)$) with the angle between two subsets of regular gradients ($\theta(R_1, R_2)$), averaged over 5 runs during honest training. The x-axis denotes the passage of time during the first training epoch.}
    \label{fig:angles}
\end{figure*}

\begin{figure*}[t!]
    \centering
    \begin{subfigure}{0.24\textwidth}
        \includegraphics[width=\textwidth]{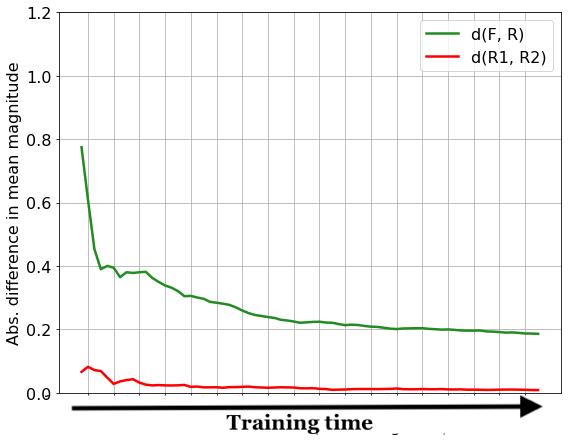}
        \caption{MNIST}
        \label{fig:mnist_mag}
    \end{subfigure}
    \begin{subfigure}{0.24\textwidth}
        \includegraphics[width=\textwidth]{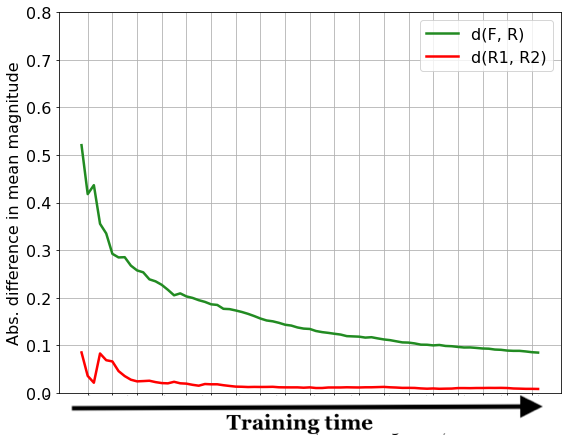}
        \caption{Fashion-MNIST}
        \label{fig:f_mnist_mag}
    \end{subfigure}
    \begin{subfigure}{0.24\textwidth}
        \includegraphics[width=\textwidth]{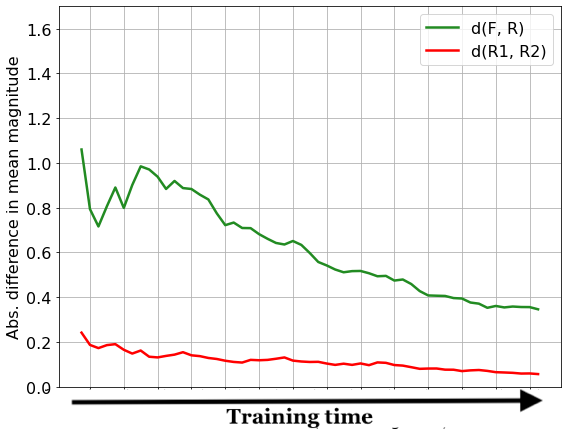}
        \caption{CIFAR10}
        \label{fig:cifar_mag}
    \end{subfigure}
    \begin{subfigure}{0.24\textwidth}
        \includegraphics[width=\textwidth]{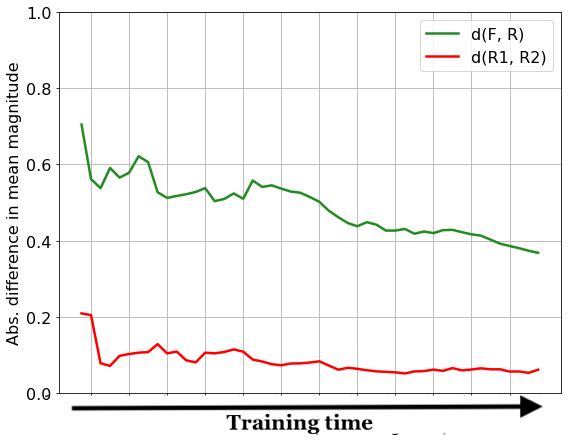}
        \caption{CIFAR100}
        \label{fig:cifar100_mag}
    \end{subfigure}
    \caption{Comparison of the average magnitude values ($d(F,R)$ and $d(R_1, R_2)$) for fake and regular gradients, averaged over 5 runs during honest training. The x-axis denotes the passage of time during the first training epoch.}
    \label{fig:mags}
\end{figure*}

We can now define the values clients compute to reach a decision. First, after each fake batch, the clients compute the value:
\begin{equation} 
    S = \frac{
                \theta(F,R) \cdot d(F,R) - \theta(R_1,R_2)\cdot d(R_1,R_2)
            }{
                d(F,R) + d(R_1, R_2) + \varepsilon
            }.
    \label{eq:main}
\end{equation}
The numerator contains the useful information we want to extract, and we divide that result by $d(F,R) + d(R_1, R_2) + \varepsilon$, where $\varepsilon$ is a small constant to avoid division by zero. This division bounds the $S$ value within the interval $[-\pi, \pi]$, a feature that will shortly come handy. 

So far, the claims lead us to consider high S values as indicating an honest server, and low S values as indicating a malicious server. However, the S values obtained during honest training vary from one model/task to another. For a more effective method, we need to define the notions of \textit{higher} and \textit{lower} more clearly. For this purpose, we will define a \textit{squashing function} that maps the interval $[-\pi, \pi]$ to the interval $(0,1)$, where high S values get mapped infinitesimally close to 1 while the lower values get mapped to considerably lower values.\footnote{From here on we will refer to the values very close to 1 as being \textit{equal} to 1, since that is the case when working with limited-precision floating point numbers.} This allows the clients to choose a threshold to separate high and low values. 

Our function of choice for the squashing function is the logistic sigmoid function $\sigma$. To provide some form of flexibility to the clients, we introduce two hyper-parameters, $\alpha$ and $\beta$, and define the function as follows:
\begin{equation}
  \tag{SplitGuard Score}
  SG = \sigma(\alpha \cdot S)^\beta \in (0,1).
\end{equation}
The function fits naturally for our purposes into the interval $[-\pi, \pi]$, mapping the high-end of the interval to 1, and the lower-end to 0. The parameter $\alpha$ determines the range of values that get mapped very close to 1, while increasing the parameter $\beta$ punishes the values that are less than 1. We discuss the process of choosing the $\alpha$ and $\beta$ in more depth in Section \ref{usage}.

\section{Results}

We need to answer five questions to claim that SplitGuard is an effective method:
\begin{itemize}
    \item How much does sending fake batches affect model performance? If the performance loss is significant, then the harm might outweigh the benefit.
    \item Do our two claims hold? 
    \item How accurately does SplitGuard detect FSHA, while not reporting an attack during honest training?
    \item What can a typical adversary learn until detection? SplitGuard's success relies on the presupposition that it can detect FSHA before the attacker achieves his goal.
    \item Can SplitGuard detect FSHA when the server includes the labels into the process as well? If not, then this is an easy way out for the attacker.
\end{itemize}
In each of the following subsections, we experimentally answer one of these questions. For our experiments, we used the ResNet architecture \cite{he2015deep}, trained with the Adam optimizer \cite{kingma_adam_2017}, on the MNIST \cite{lecun2010mnist}, Fashion-MNIST \cite{xiao2017/online}, and CIFAR10/100 \cite{Krizhevsky09learningmultiple} datasets. We implemented our attack in Python (v 3.7) using the PyTorch library (v 1.9) \cite{pytorch}. In all our experiments, we limit our scope only to the first epoch of training. It is the least favorable time for detecting an attack since the model initially behaves randomly, and represents a lower bound for results in later epochs.

\begin{table}[h!]
    \centering
    \caption{Test classification accuracy values of the ResNet model for the MNIST, F-MNIST, and CIFAR10/100 datasets for different $B_F$ values after three epochs of  training with SplitGuard, averaged over 10 runs with a $P_F$ of $0.1$.}
    \label{tab:accuracies}
    
    \begin{tabular}{ccccc}
    \toprule
        $B_F$ & \multicolumn{4}{c}{Classification Accuracy (\%)} \\ \midrule
                           & \textbf{MNIST} & \textbf{F-MNIST} & \textbf{CIFAR10} & \textbf{CIFAR100}\\ \midrule
        $0$ (Original)     & 98.68 & 89.60 & 64.24 & 36.5 \\ \midrule
        $8/64$             & 98.62 & 89.80 & 64.0 & 38.46 \\ 
        $16/64$            & 98.86 & 89.24 & 66.38 & 38.24 \\ 
        $32/64$            & 99.14 & 90.40 & 62.58 & 36.06 \\ 
        $64/64$            & 98.92 & 89.44 & 63.28 & 35.86 \\ 
    \bottomrule
    \end{tabular}
\end{table}

\subsection{Effect on Model Performance}

Table \ref{tab:accuracies} displays the classification accuracy of the ResNet model on the test sets of our four benchmark datasets with different $B_F$ values after three epochs of training, averaged over 10 runs. The client model consists of a single convolutional layer, and the rest of the model is computed by the server. This is the worst-case scenario for this purpose, since the part of the model that is being updated with fake batches is as large as possible.

The results show that the model performs similarly when trained with and without SplitGuard. There is not a noticeable and consistent decrease in performance for any of the datasets, even for high $B_F$ values such as 1. 

\subsection{Validating the Claims} \label{validation}

Going back to our two claims, we now demonstrate that fake gradients make a larger angle with regular gradients than the angle between two random subsets of regular gradients, and that fake gradients have a higher magnitude than regular gradients. For each dataset, Figures \ref{fig:angles} and \ref{fig:mags} display these values obtained during the first epoch of training with an honest server, averaged over 5 runs. 

Figure \ref{fig:angles} shows that $\theta(F,R)$ is consistently greater than $\theta(R_1, R_2)$. Note however that the difference is greater for MNIST (around $60^{\circ}$) than for Fashion-MNIST (around $30^{\circ}$) and the CIFAR datasets (around $10^{\circ}$). This means that as the model performs better (see Table \ref{tab:accuracies}), the difference between the angles becomes higher as well. 

Figure \ref{fig:mags} displays a similar relation between the $d(F,R)$ and $d(R_1, R_2)$ values obtained during the first epoch of training. For each of our datasets, $d(F,R)$ values are consistently higher than the $d(R_1, R_2)$ values, although the difference is smaller for CIFAR compared to MNIST for similar reasons with the angle values.

\begin{figure*}[t!]
    \centering
    \begin{subfigure}{0.24\textwidth}
        \includegraphics[width=\textwidth]{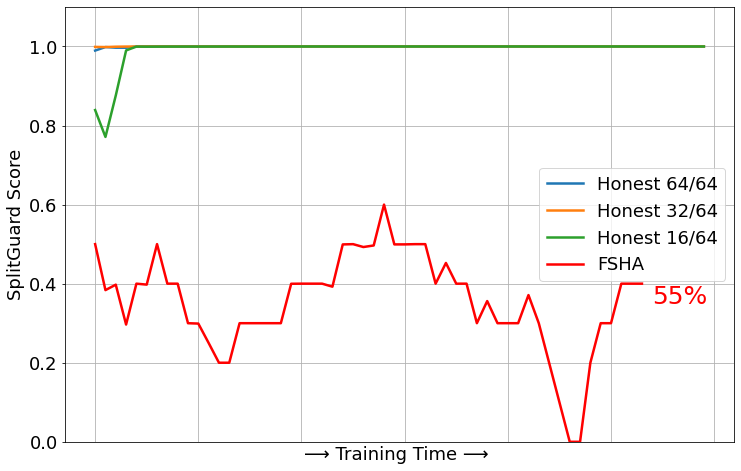}
        \caption{MNIST}
        \label{fig:mnist_results}
    \end{subfigure}
    \begin{subfigure}{0.24\textwidth}
        \includegraphics[width=\textwidth]{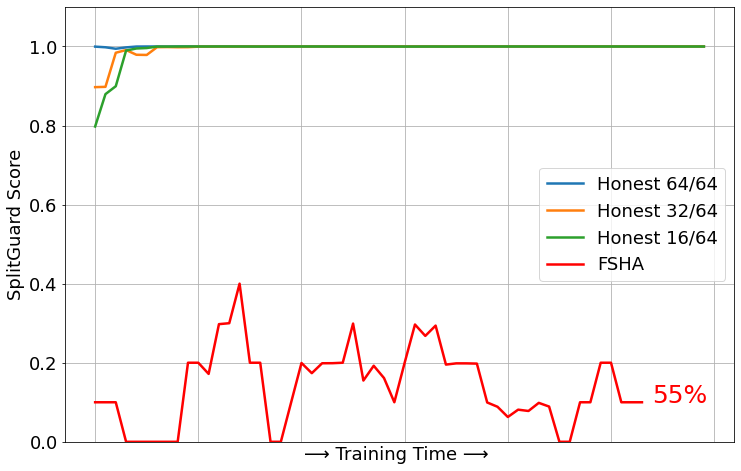}
        \caption{Fashion-MNIST}
        \label{fig:f_mnist_results}
    \end{subfigure}
    \begin{subfigure}{0.24\textwidth}
        \includegraphics[width=\textwidth]{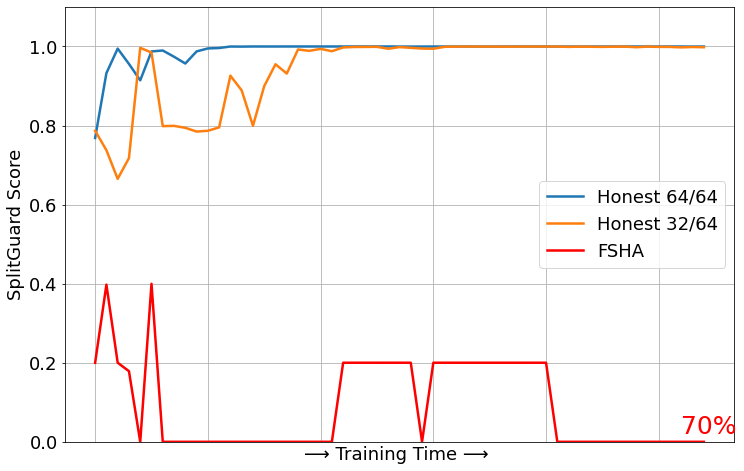}
        \caption{CIFAR10}
        \label{fig:cifar_results}
    \end{subfigure}
    \begin{subfigure}{0.25\textwidth}
        \includegraphics[width=\textwidth]{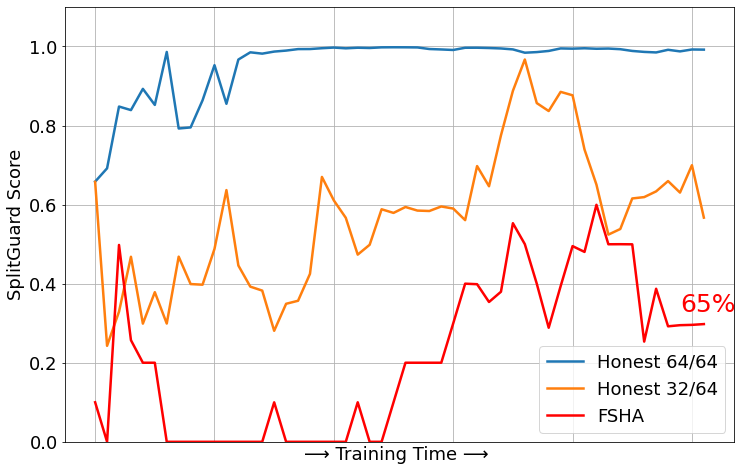}
        \caption{CIFAR100}
        \label{fig:cifar100_results}
    \end{subfigure}
    \caption{SplitGuard scores obtained while training with an honest server, and a FSHA attacker until detection during the first epoch, averaged over 5 runs. The x-axis displays the flow of time during the first epoch. The $P_F$ value is set to $0.1$, and the $B_F$ values vary when training with an honest server. Lower $B_F$ values are excluded from the CIFAR10/100 results since they do not serve a purpose given the model's classification accuracy (see Section \ref{decision_making}). The labels at the end of FSHA lines correspond to the average detection times as the amount of progress within the first epoch, as shown in Table \ref{tab:detection_results} as well.}
    \label{fig:main_results}
\end{figure*}

To recap, Figures \ref{fig:angles} and \ref{fig:mags} demonstrate that our claims are valid during the first epoch of training for our benchmark datasets. The decreasing difference as the models become less adept (going from MNIST to CIFAR100) implies that the protocol might need to be extended beyond the first epoch for more complex tasks.

\subsection{Detecting FSHA} \label{detecting_fsha}

With the claims validated, the questions of actual effectiveness remains: how accurately can SplitGuard detect FSHA?

\subsubsection{Distinguishable Scores}

\begin{table*}[t!]
    \centering
     \caption{Attack detection statistics for the five example policies, collected over 100 runs of the first epoch of training with a FSHA attacker and an honest server. The true positive rate (TPR) corresponds to the rate at which SplitGuard succeeds in detecting FSHA. The false positive rate (FPR) corresponds to the share of honest training runs in which SplitGuard mistakenly reports an attack. The $t$ field denotes the average point of detection (as the share of total batches).}
    \label{tab:detection_results}
    \begin{tabular}{l|ccc|ccc|ccc|ccc}
    \toprule
        \textbf{Policy} & \multicolumn{3}{c}{\textbf{MNIST}} & \multicolumn{3}{|c|}{\textbf{F-MNIST}} & \multicolumn{3}{c|}{\textbf{CIFAR10}} & \multicolumn{3}{c}{\textbf{CIFAR100}} \\ \midrule
                    & TPR & FPR   & $t$    & TPR & FPR & $t$ & TPR & FPR & $t$ & TPR & FPR & $t$  \\ \midrule
        Fast        & 1   & 0.01  & 0.016  & 1  & 0.09 & 0.016  & 1  & 0.20 & 0.11  & 1  & 0.90 & 0.12  \\
        Avg-10      & 1   & 0     & 0.14   & 1  & 0.03 & 0.14  & 1  & 0.29 & 0.20 & 1  & 0.75 & 0.14  \\ 
        Avg-20      & 1   & 0     & 0.24   & 1  & 0.01 & 0.24  & 1  & 0.21 & 0.33 & 1  & 0.45 & 0.26  \\ 
        V      & 1   & 0     & 0.55   & 1  & 0    & 0.55  & 1  & 0.02 & 0.70 & 1  & 0.11 & 0.65  \\
    \bottomrule
    \end{tabular}
   
\end{table*}

Figure \ref{fig:main_results} compares the SplitGuard scores obtained against a malicious (FSHA) server and an honest server, averaged over 5 runs with a $P_F$ value of 0.1 and varying $B_F$ values.\footnote{The $B_F$ values do not affect the SplitGuard scores obtained against a FSHA server, since the client's loss function $L_f$ is independent of the labels, though as we will discuss later there might be strategic reasons for choosing different $B_F$ values.} We set the $\alpha$ and $\beta$ values to 7 and 1 for all datasets; we discuss further in Section \ref{usage} the way the $\alpha$ and $\beta$ values can be chosen.

The results displayed in Figure \ref{fig:main_results} indicate that the SplitGuard scores are distinguishable enough to enable detection by the client. The SplitGuard scores obtained against an honest server are very close or equal to 1, while the scores obtained against a FSHA server do not surpass $0.8$, and vary more vigorously. Higher $B_F$ values are more effective. For example, it takes slightly more time for the scores to get fixed around 1 for Fashion-MNIST with a $B_F$ of $4/64$ compared to a $B_F$ of 1. 

\subsubsection{Decision Policies}

To assess more rigorously how accurate SplitGuard is at detecting FSHA, and likewise not reporting an attack during honest training, we define three candidate \textbf{decision policies} with different goals and test each one's effectiveness. A policy takes as input the list of SplitGuard scores obtained up to that point, and decides if the server is launching a training-hijacking attack or not. 

We set a threshold score of 0.9 for these example policies. While the clients can choose different thresholds (Section \ref{decision_making}), the results in Figure \ref{fig:main_results} indicate that 0.9 is a reasonable starting point. The three policies, also displayed in Algorithm \ref{alg:policies} are defined as follows:
\begin{itemize}
    \item \textit{Fast:} Fix an early batch index. Report an attack if the last score obtained is less than $0.9$ after that index. The goal of this policy is to detect an attack as fast as possible, without worrying about a high false positive rate.
    \item \textit{Avg-$k$:} Report an attack if the average of the last $k$ scores is less than $0.9$. This policy represents a middle point  between the \textit{Fast} and the \textit{Voting} policies. 
    \item \textit{Voting:} Divide the scores sequentially into groups of a fixed size and calculate each group's average. Report attack if the majority of the means is less than $0.9$. This policy aims for a high overall success rate (i.e. high true positive and low false positive rates); it can tolerate making decisions relatively later. 
\end{itemize}
We will discuss the clients' decision-making process beyond these sample policies in more detail in Section \ref{decision_making}.

\begin{algorithm}[t!]
\SetKwFunction{fast}{FAST}
\SetKwFunction{avg}{AVG-K}
\SetKwFunction{voting}{VOTING}
\SetKwProg{Fn}{Function}{:}{}

\Fn{\fast{$S$: scores, $T$: threshold}}{
    \Return $S[-1] < T$
}

\Fn{\avg{$S$: scores,  $k$: number of scores, $T$: threshold}}{
    \Return \texttt{mean}$(S[-k:]) < T$
}

\Fn{\voting{$S$: scores, $n$: group size, $T$: threshold}}{
    votes = 0\\
    c = $\lceil len(scores) / n \rceil $ // group count \\
    // default $c=10$ and $n=5$ \\
    \For{$i$ from $0$ to $c$}{
        group = S[$i \cdot n$ : $(i+1) \cdot n$] \\
        \uIf{\texttt{mean}(group) $< T$}{
            votes += 1 \\
        }
    }
    \Return $votes > c/2$ 
}

\caption{Example Detection Policies}
\label{alg:policies}
\end{algorithm}

\begin{figure*}[t!]
    \centering
    \includegraphics[width=\textwidth]{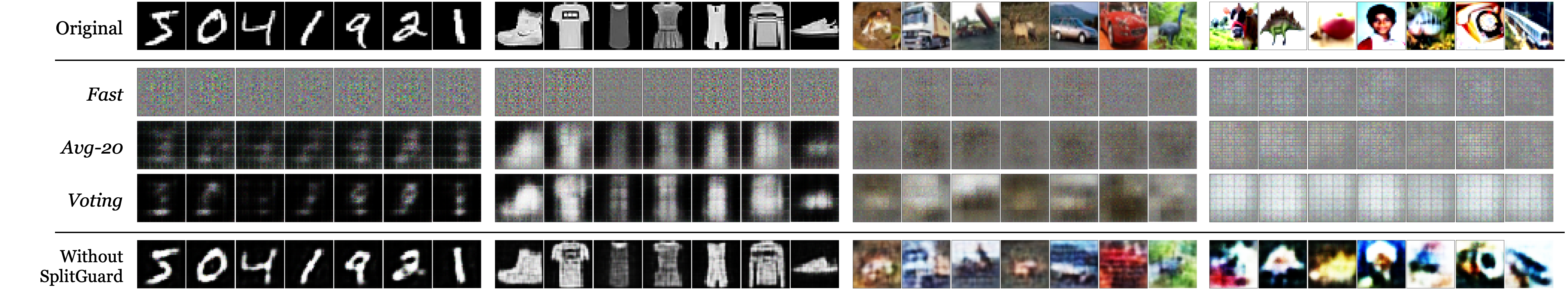}
    \caption{Results obtained by a FSHA attacker for the MNIST, F-MNIST, CIFAR10, and CIFAR100 datasets until the average detection times of the given policies as displayed in Table \ref{tab:detection_results}. The first row displays the original images, and the last row displays the results obtained by a FSHA attacker able to run for an arbitrary duration without being detected.}
    \label{fig:fsha_results}
\end{figure*}

Table \ref{tab:detection_results} displays the detection statistics for each of these strategies obtained over 100 runs of the first epoch of training against a FSHA attacker and an honest server with a $B_F$ of 1 and $P_F$ of 0.1. For the \textit{Avg-k} policy, we use $k$ values of 10 and 20; this ensures that the policy can run within the first training epoch.\footnote{With a batch size of 64, one epoch is equal to 938 batches for MNIST and F-MNIST, and 782 for CIFAR10/100.} For the \textit{Voting} policy, we set the group size to 5. Finally, we set $N$, the index at which SplitGuard starts running, as 20 for MNIST and F-MNIST, 50 for CIFAR10, and 100 for CIFAR100.\footnote{The models initially behave randomly. We want to exclude those periods from SplitGuard.}

Most significantly, all the strategies achieve a perfect true positive rate (i.e. successfully detect all runs of FSHA). Expectedly, the \textit{Fast} strategy achieves the fastest detection times as denoted by the $i$ values in Table \ref{tab:detection_results}, detecting in at most a hundred training batches all instances of the attack. 

False positive rates increase as the model's performance decreases, moving from MNIST to F-MNIST and then to CIFAR10/100. This means that more training time should be taken to achieve higher success rates in more complex tasks. However, as we will observe in Section \ref{sec:fsha_result}, the model not having a high performance also implies that FSHA will be less effective. Nevertheless, the \textit{Voting} policy achieves a false positive rate of 0 for (F-)MNIST, 0.02 for CIFAR10, and 0.11 for CIFAR100, indicating that despite the relatively high false positive rates of the \textit{Fast} and \textit{Avg-$k$} policies, better detection performance in less time is achievable through more sophisticated policies, such as the \textit{Voting} policy.

\subsection{What Does the Attacker Learn Until Detection?} \label{sec:fsha_result}

We now analyze what an FSHA adversary can learn until the detection batch indices displayed in Table \ref{tab:detection_results}. Figure \ref{fig:fsha_results} displays the results obtained by the attacker when it runs the attack only until the detection batch indices for the given policies, and compares them with the attacker's results in the no-SplitGuard scenario. 

For the \textit{Fast} policy, the attacker obtains not much more than random noise; if a high false positive rate can be tolerated (e.g. privacy of the data is highly critical, and the server is distrusted), this policy can be applied to prevent any data leakage.

The attack results get more accurate as the attacker is given more time. Nevertheless, especially for the more complex CIFAR10/100 tasks, the results obtained by the attacker against the \textit{Voting} policy do not contain the distinguishing features of the original images, whereas the no-SplitGuard scenario produces results that are visually more accurate. This highlights the effectiveness of the \textit{Voting} policy, preventing significant information leakage with a relatively lower false positive rates. 

Finally, the CIFAR10/100 results also show that the attacker having more time for a more complex task is tolerable because after the same number of batches, the attacker's results for MNIST and Fashion-MNIST are more accurate compared to the CIFAR10/100 results.

\begin{figure}[h!]
    \centering
    \includegraphics[width=\linewidth]{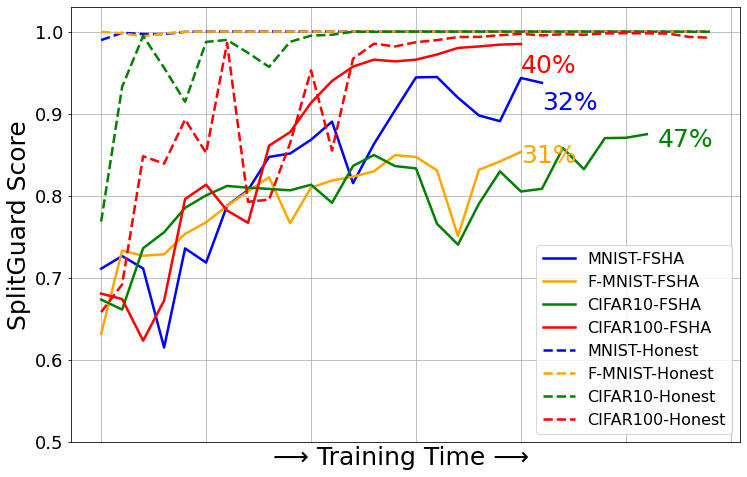}
    \caption{SplitGuard scores for our four benchmark datasets against an honest server anda FSHA server performing multitask learning, averaged over 5 runs with $P_F=0.1$ and $B_F=1$. The x-axis represents the passage of training time. The labels at the end of each line displays the average progress within the first epoch when SplitGuard detects FSHA using the Voting decision policy.}
    \label{fig:multitask}
\end{figure}

\begin{figure*}[t!]
    \centering
    \includegraphics[width=\textwidth]{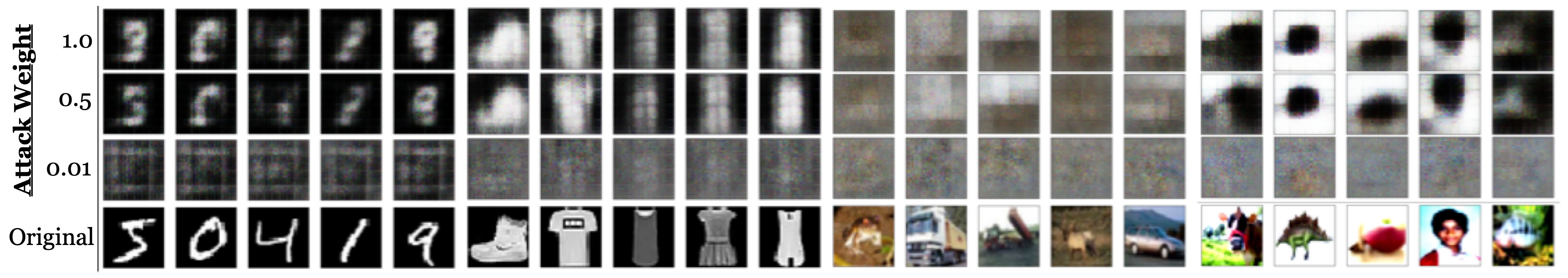}
    \caption{Results the FSHA attacker obtains when it performs multitask learning until detection for the MNIST, F-MNIST, CIFAR10, and CIFAR100 datasets, with 5 images randomly selected from each dataset. The bottom row displays the original inputs, and the top three rows display the attack's results when the attack weight is 1, 0.5, and 0.01. }
    \label{fig:mt_fsha_results}
\end{figure*}

\begin{table}[t!]
    \centering
     \caption{Detection results on MNIST, F-MNIST, and CIFAR10/100 datasets for our four policies against a server launching FSHA while performing multitask learning, averaged over 100 runs. The TPR column displays the true positive (i.e. success) rate and the $t$ values correspond to the average share of total batches trained on until detection. Detecting at later batches gives the attacker more time while decreasing the false positive rate (see Table \ref{tab:detection_results}); all the displayed batch indices fall within the first epoch. There is no false positive rate column since they only matter for the honest training scenario which is displayed in Table \ref{tab:detection_results}.}
    \label{tab:multitask_detection_results}
    \begin{tabular}{l|cc|cc|cc|cc}
    \toprule
        \textbf{Policy} & \multicolumn{2}{c}{\textbf{MNIST}} & \multicolumn{2}{|c|}{\textbf{F-MNIST}} & \multicolumn{2}{c|}{\textbf{CIFAR10}} & \multicolumn{2}{c}{\textbf{CIFAR100}} \\ \midrule
                    & TPR & $t$ & TPR & $t$ & TPR & $t$ & TPR & $t$  \\ \midrule
        Fast        & 1   & 0.015  & 1   & 0.028  & 1   & 0.03  & 1   & 0.14  \\
        Avg-10      & 1   & 0.11 & 1   & 0.11 & 1   & 0.14 & 1   & 0.14 \\ 
        Avg-20      & 1   & 0.22 & 1   & 0.22 & 1   & 0.26 & 1   & 0.26  \\ 
        Voting      & 1   & 0.32 & 1   & 0.31 & 1   & 0.47 & 1   & 0.40  \\
    \bottomrule
    \end{tabular}
   
\end{table}

\subsection{Multitask Learning}

As a response to the preceding discussion, the question might arise of the server somehow including the label values in the attack in an attempt to subvert the detection process.\footnote{This only concerns the shared-label SplitNN setup (Figure \ref{fig:splitnn_label_sharing}) since in the private-label scenario the server does not have access to the label values input to the classification loss.} A reasonable way of doing this is to make the client optimize both the FSHA loss and the classification loss functions, e.g. by computing their weighted average, an \textit{attack weight} of 1 meaning plain FSHA and 0 no attack.

Figure \ref{fig:multitask} displays the SplitGuard scores obtained against a malicious server returning the average of FSHA and classification losses back to the client, averaged over 5 runs with a $P_F$ of 0.1 and $B_F$ of 1. 

The attacker performing multitask learning gives rise to higher SplitGuard scores than those displayed in Figure \ref{fig:main_results} as the randomized labels have \textit{some} influence over the parameter updates of the client model. The scores also tend to generally increase with time; since the fake/regular batch discrepancy becomes stronger as the model learns to classify more accurately. However, the scores are still noticeably smaller than 1 during detection time, indicating that the server performing multitask learning is not enough to yield SplitGuard ineffective. 

Further quantifying the effect of multitask learning, Table \ref{tab:multitask_detection_results} displays the rates at which our three sample policies detect a FSHA server performing multitask learning (true positive rate). As in the original scenario (Table \ref{tab:detection_results}), all policies detect all instances of the attack. These results confirm the visual observation we made in the previous paragraph that the server performing multitask learning does not yield SplitGuard ineffective. 

Finally, Figure \ref{fig:mt_fsha_results} displays the results obtained by a FSHA server performing multitask learning with varying attack weights. A very low attack weight such as 0.01 produces random-looking results, while the distinction between a full- and half-attack result is less clear. However, it is not clear whether this is an inherent property of the attack, or is specific to the dataset or client/server model architectures. 

\section{Discussion} \label{usage}

\begin{table}[h!]
    \caption{Classification accuracies (averaged over 5 runs) over the test set for the original model and the client's linear classifier for honest and random training scenarios. The linear classifier was introduced with the idea that perhaps its low performance could alert the client to an attack.}
    \centering
    \begin{tabular}{lcccc}
    \toprule
          & \multicolumn{4}{c}{Classification Accuracy (\%)} \\ \midrule
        \textbf{Training}  & \multicolumn{2}{c}{Honest} & \multicolumn{2}{c}{Random} \\ \midrule
        \textbf{Model}     & Original & Linear           & Original & Linear \\ \midrule
        MNIST              & 98.09    & 92.13   & 8.62     & 91.54 \\
        F-MNIST            & 86.13    & 81.53   & 9.40     & 83.18 \\
        CIFAR10            & 51.62    & 28.26   & 8.08     & 28.88 \\
        CIFAR100           & 22.92    & 9.26    & 0.94     & 8.86 \\
    \bottomrule
    \end{tabular}

    \label{tab:linear_classifier}
\end{table}

\subsection{Training-Hijacking and Learning a Classifier from the Clients' Outputs}

A reasonable argument against a training-hijacking detection protocol such as SplitGuard is that it might have been simpler (and practically impossible for the attacker to detect) if the clients, already knowing the label values, had attempted to train a classifier using their models' outputs and assessed its accuracy, the hypothesis being that training-hijacking would lead the clients towards outputting values from which it would be difficult to a train a classifier, whereas the values output by an honestly-trained client model would by necessity give rise to an accurate classifier. However, the weak point of the argument is clear: The values clients output against a training-hijacking server \textit{must} contain significant information (and this necessarily includes the information that would help classify them) about the inputs so that the attacker can recover the inputs from the outputs.

We experimentally tested this argument for the worst-case scenario: The server learns random labels (effectively leading the clients towards outputting random values, rather than the FSHA server attempting to extract as much information as possible), and the client's classifier consists of a single linear layer. 
Table \ref{tab:linear_classifier} displays the original model's and the linear classifier's classification accuracies for the honest and random training scenarios for each of our four benchmark datasets averaged over five runs. In honest training, the linear classifier performs worse than the original, much more sophisticated, model. But most significantly, there is no noticeable difference between the linear classifier's performances in honest- and random-training scenarios; the difference is at most $1-2\%$, sometimes in the favor of the random-training linear classifer (see F-MNIST and CIFAR10 in Table \ref{tab:linear_classifier}). 

In short, attempting to train a linear classifier using the client's outputs does not provide a reliable detection mechanism against training-hijacking attacks. 

\subsection{Computational Complexity}

SplitGuard does not incur a significant computational cost regarding time or space. 

Since SplitNN clients are already assumed to be able to run back-propagation on a few DNN layers, calculating the S value described in Equation \ref{eq:main} is a simple task. Quantitavely, averaged over 10 runs, an epoch of training in our experimental setup with CIFAR-10 takes 24.16 seconds with, and 19.73 seconds without SplitGuard. 

Space-wise, although it might seem like storing the gradient vectors for potentially multiple epochs requires a significant amount of space, the clients do not have to store all the gradient vectors. For each of the sets $F$, $R_1$, $R_2$, the clients have to maintain two quantities: a sum of all vectors in the set, and the average magnitude of the vectors in the set; the first has the dimensions of a single gradient vector, and the second is a scalar. More importantly, both of these quantities can be maintained in a running manner. This keeps the total space required by SplitGuard to $O(1)$ with respect to training time, equivalent to the space needed for three scalar values and three gradient vectors. For reference, the space required to store a single gradient vector in our experiments was 2.304 KB. Since the space requirement is independent of the total number of batches, it is possible to run SplitGuard during arbitrarily long training processes.

\begin{algorithm}[h!]

\begin{enumerate}
    \item Choose parameters $\alpha, \beta, B_F, P_F, N, T$ (Section \ref{section_prep}); simulating different server behaviors if possible. 

    \item Choose a decision policy based on user goals \\(Section \ref{detecting_fsha}). 

    \item Start training and evaluate scores after each fake batch (Algorithm \ref{alg:main}, Section \ref{section_training}).
    
    \item Stop training if server is likely attacking.
\end{enumerate}

\caption{End-to-End SplitGuard Outline}
\label{alg:endtoend}
\end{algorithm}

\subsection{Using SplitGuard} \label{decision_making}

There are two steps to using SplitGuard: preparation before training, and actual use during training. Algorithm \ref{alg:endtoend} provides an outline of the end-to-end procedure and we further detail the two parts below.

\subsubsection{Preparation Before Training} \label{section_prep}

Before training, SplitGuard's various parameters should be set. For choosing $\alpha$ and $\beta$ values, we consider two scenarios: the clients know or do not know the model architecture.

If the clients know the server-side model architecture, then they can simulate different server behaviors using (part of) their local data. For example, they can train the entire model against simulated honest and random-labeling servers. They can then set the parameters $\alpha$ and $\beta$ to map the S values (Equation \ref{eq:main}) obtained against the honest server close to 1, and those obtained against a random-labeling server close to 0. In this scenario, since the clients' confidence on the accuracy of the method is expected to be higher, a relatively high threshold can be set, such as $0.95$.

If the clients do not know the model architecture, then they should set the parameters $\alpha$ and $\beta$ in a more ad hoc manner. Nevertheless, S values all lying within the interval $[-\pi, \pi]$ makes the clients' job easier. It is unreasonable to set extremely high $\alpha$ or $\beta$ values since they will cause the squashing function to make sudden jumps, or map no value close to one. Lower threshold values, or protocols with later decision points can be utilized to compensate for this uncertainty. Nevertheless, as our experiments also demonstrate, smaller values such as $\alpha=7$ and $\beta = 1$ are reasonable starting points.

Once the $\alpha$ and $\beta$ values are set, we can consider the other parameters ($B_F$, $P_F$, and $N$). Each parameter involves a different trade-off:
\begin{itemize}
    \item \textbf{Probability of sending  a fake batch} ($P_F$). 
        \begin{itemize}
            \item ($+$) Higher $P_F$ values mean more fake batches, and thus a more representative sample of fake gradient values, increasing the effectiveness of the method. 
            \item ($-$) Higher $P_F$ values can also degrade model performance, since the server model will be learning random labels for a higher number of examples, and a higher share of the potentially scarce dataset will be allocated for SplitGuard.
        \end{itemize} 
    \item \textbf{Number of randomized labels in each batch} ($B_F$). 
        \begin{itemize}
            \item ($+$) More random labels in a batch means that fake batches and regular batches behave even more differently, and the method becomes more effective.
            \item ($-$) Depending on the model's training performance, batches with entirely random labels can be detected by the server. One way to overcome this difficulty is to perform the loss computation on the client side.  
        \end{itemize} 
    \item \textbf{Number of initial batches to ignore} ($N$).
        \begin{itemize}
            \item ($+$) A smaller $N$ value means that the server's malicious behavior can be detected earlier, giving it less time to attack.
            \item ($-$) Since a model behaves randomly in the beginning of the training, the initial batches are of little value for our purposes. Computing SG scores for later batches will make it easier to distinguish honest behavior, but in return give the attacker more time.
        \end{itemize} 
\end{itemize}

\subsubsection{During Training} \label{section_training}

After each fake batch, clients can make a decision on whether the server is launching an attack or not. The main decision procedure is as follows:
\begin{outline}[enumerate]
    \1 Is the SG value \textbf{high or low}?
        \2 If \textbf{high}, there are no problems. Keep training.
        \2 If \textbf{low}, there are two possible explanations:
            \3 The model has not learned enough yet. Keep training, potentially making changes. 
            \3 The server is launching an attack. Halt training.
\end{outline}
The policies in Section \ref{detecting_fsha} did not consider the first explanation (1.b.i) of low scores, namely the model not having learned enough. Taking that into consideration could help reduce the false positive rates since the main reason behind false positives was the model's random initial behavior. 

\textbf{Explaining Low Scores.} When a client decides that the SplitGuard score is low, it should choose between two possible explanations: either the model has not learned enough yet, or the server is launching a training-hijacking attack. 

Informally, a low score indicates that fake gradients are similar to regular gradients; i.e. the model behaves similarly when given fake batches and regular batches. For classification, \textit{behaving similarly} is equivalent to having a similar classification accuracy. Then, the explanation that the model has not learned enough yet is more likely if the expected classification accuracy for a fake batch is close to the actual (expected) prediction accuracy. If these values are different but the SplitGuard score is still low, then the server is very likely launching an attack.

\begin{figure}[t!]
    \centering
    \includegraphics[width=0.5\textwidth]{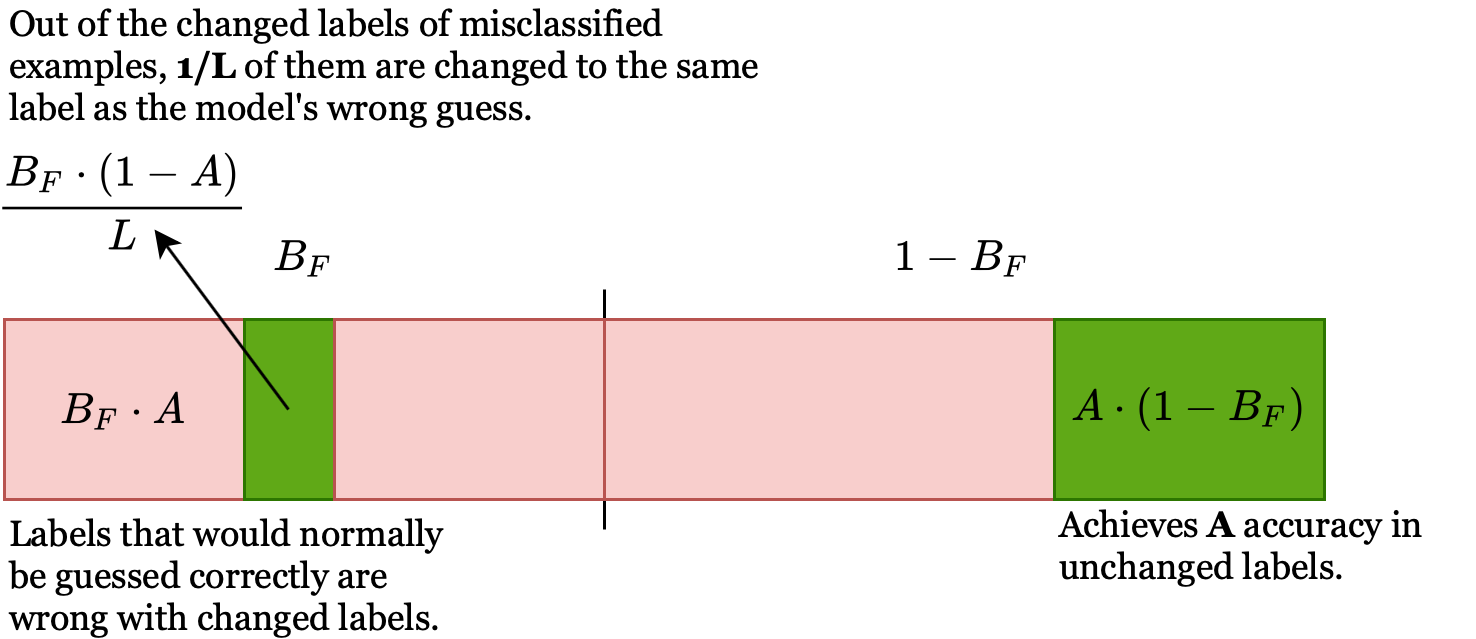}
    \caption{Expected classification accuracy in a fake batch with the share $B_F$ of the labels randomized. The model normally has classification accuracy $A$ with $L$ labels.}
    \label{fig:accuracy}
\end{figure}

We can formulate the expected accuracy for a fake batch. Say the total number of labels is $L \in \mathbb{N}\ (L \geq 2)$ and the overall model has classification accuracy $A \in [1/L, 1]$. Then the expected classification accuracy for a fake batch with the share $B_F \in [0,1]$ of the labels randomized is
\begin{equation}
    A_F = A \cdot (1-B_F) + \frac{B_F \cdot (1-A)}{L}.
    \label{eq:accuracy}
\end{equation}
Figure \ref{fig:accuracy} explains this equation visually.

If the model terminates on the client-side (as in Figure \ref{fig:splitnn_private_labels}), then the clients already know the exact accuracy value. If that is not the case but the clients know the model architecture on the server side, then they can train the model using their local data, and obtain an estimate of the expected classification accuracy of the actual model during the first epoch. If even that is not possible, then in the worst-case the clients can train a linear classifier appended to their model to obtain a lower bound on the original model accuracy.\footnote{A related, interesting study concludes that what a neural network learns during its initial epoch of training can be \textit{explained} by a linear classifier \cite{kalimeris_sgd_2019}, in the sense that if we know the linear model's output, then knowing the main model's output provides almost no benefit in predicting the label. Note however that this does not hold for \textit{any} linear classifier, but the optimal one.}

Formalizing this discussion, for SplitGuard to be effective, it must be the case that $A >> A_F$. If $A_F \approx A$, then the clients' choice of $B_F$ is not right, and they should increase it. $A_F$ is a linear function of $B_F$ with the coefficient 
$$
-A + \frac{1}{L} - \frac{A}{L}.
$$
Since $A \in [1/L, 1]$,
$$
-A + \frac{1}{L} \leq 0
$$
and 
$$
-A + \frac{1}{L} - \frac{A}{L} \leq 0
$$
as well. Thus, $A_F$ is indeed a monotonic function of $B_F$, and increasing $B_F$ either keeps $A_F$ constant or decreases it. Then when the clients decide that the SG value is low and that $A_F \approx A$, the best course of action is to increase $B_F$. If $B_F$ is already 1, then clients should wait until the model becomes sufficiently accurate so that a completely randomized batch makes a difference. As discussed previously, this is not a worrisome scenario, since the attack's effectiveness also relies on the model's adeptness.

\begin{algorithm}[h!]
\SetKwFunction{func}{MAKE\_DECISION}
\SetKwProg{Fn}{Function}{:}{}

$A$: Model's classification accuracy \\
$A_F$: Expected classification accuracy for a fake batch \\
$B_F$: Share of randomized labels in a fake batch  \\
$N$: Number of initial batches to ignore \\
\Fn{\func{$F, R$}}{
    \uIf{scores are high}{ 
        Keep training.
    } 
    \uElseIf{$A \approx A_F$}{
        \uIf{$B_F = 1$}{
            It is too early to detect. Wait.
        } \Else{
            Increase $B_F$.
        }
        [Optional] Increase $N$.
    }
    \uElse{
        The server is launching an attack. Stop training.
    }
}
\caption{Clients' decision-making process}
\label{alg:decision}
\end{algorithm}

Finally, an alternative course of action is to increase $N$, discarding the initial group of gradients. Since the models behave randomly in the beginning, increasing $N$ decreases the noise, and can help distinguish an honest server from a malicious one. Increasing $N$ is a reversible process, provided that clients store the gradient values.

With these discussions, we can finalize the clients' decision-making process as the function \texttt{MAKE\_DECISION}, displayed in Algorithm \ref{alg:decision}.

\subsection{Detection by the Attacker} \label{detection}

An attacker can in turn try to detect that a client is running SplitGuard. It can then try to circumvent SplitGuard by using a legitimate surrogate model learning the original classification task.

If the server controls the model's output (Figure \ref{fig:splitnn_label_sharing}), then it can detect if the classification error of a batch is significantly higher than the other ones. Since SplitGuard is a potential, though not the only, explanation of such behavior, this presents an opportunity for an attacker to detect it. However, the model behaving significantly differently for fake and regular batches also implies that the model is at a stage at which SplitGuard is effective. This leads to an interesting scenario: since the attack's and SplitGuard's effectiveness both depends on the model learning enough it seems as if the attack cannot be detected without the attacker detecting SplitGuard and vice versa.\footnote{Our claim with the preceding presentation however was that the discrepancy between fake and regular gradients precedes the FSHA server being able to extract useful information.}

We argue that this is not the case, due to the clients being in charge of setting the $B_F$ value. For example, with the MNIST dataset for which the model obtains a classification accuracy around $98\%$ after the first epoch of training, a $B_F$ value of $4/64$ results in an expected classification accuracy of $91.8\%$ for fake batches (Equation \ref{eq:accuracy}). The SplitGuard scores on the other hand displayed in Figure \ref{fig:main_results} being very close to one implies that an attack can be detected with such a $B_F$ value. Thus, clients can make it difficult for an attacker to detect SplitGuard by setting the $B_F$ value more smartly, rather than setting it blindly to 1.

Finally, we strongly recommend that a secure SplitNN setup follow the three-part setup shown in Figure \ref{fig:splitnn_private_labels} to prevent the clients sharing their labels with the server. This way, an attacker would not be able to see the accuracy of the model, and it would become harder for it to detect SplitGuard. 

\subsection{Generalizing SplitGuard}

In the form we have discussed so far, a question might arise regarding SplitGuard's effectiveness in different learning setups. We argue however that since the claims (namely that \textit{the model's behavior on the parameter space varies noticeably between two opposite tasks}) underlying SplitGuard are applicable to any kind of neural network learning on any kind of data, SplitGuard is generalizable to different data modalities, or more complex architectures. 

Note that the input modality (image, text, numeric etc.) does not take part in our presentation of SplitGuard, although we limit ourselves to the image domain since the FSHA is also restricted as such. The data modality does not affect the clients' ability to manipulate the label values as required by SplitGuard. 

Another direction of generalization is towards different attacks. Although there are no training-hijacking attacks other than FSHA against which we can test SplitGuard, we claim that SplitGuard can generalize to future attacks as well. After all, SplitGuard relies only on the assumption that randomizing the labels affects an honest model more than it affects a malicious model. Thus, to go undetected by SplitGuard, an attack should either involve learning significant information about the original task, which would likely reduce the attack's effectiveness, or craft a different loss function for each label, which could easily be prevented by not sharing the labels with the server (Figure \ref{fig:splitnn_private_labels}).

Finally, SplitGuard also generalizes to multiple-client SplitNN settings. Each client can independently run SplitGuard, with their own choices of parameters. Each client would then be making a decision regarding its own training process. Alternatively, if the clients trust each other, they can choose one client to run SplitGuard in order to minimize its effect on performance loss, or they can combine their collected gradient values and reach a collective decision. 

\subsection{Use Cases}

We now describe three potential real-world use cases for SplitGuard, modeling clients with different capabilities at each scenario.

\textbf{Powerful Clients.} A group of healthcare providers decide to train a DNN using their aggregate data while maintaining data privacy. They decide on a training setup, and establish a central server.\footnote{Alternatively, members of the group can take turns acting as the SplitNN server in a P2P manner.} Each client knows the model architecture and the hyperparameters, and preferably has access to the model's output as well (no label-sharing). The clients can train models using their local data to determine the parameters $\alpha$ and $\beta$. Each client can then run SplitGuard during their training turns and see if they are being attacked. This is an example scenario with the clients as powerful as possible, and thus represents the optimal scenario for running SplitGuard.

\textbf{Intermediate Clients.} The SplitNN server is a researcher, attempting to perform privacy-preserving machine learning on some private dataset of some data-holder (the client). The researcher designs the training procedure, but the data-holder actively takes part in the protocol. The data-holder thus has tight control over how its data is organized. The client cannot train a local model since it does not know the entire architecture, and should set the parameters $\alpha$ and $\beta$ manually. Nevertheless, it can easily run SplitGuard by modifying the training data being used in the protocol.

\textbf{Weak Clients.} An application developer is the SplitNN server, and the users' mobile devices are the clients with private data. The clients do not know the model architecture, and cannot manipulate how their data is shared with the server. The application developer is in control of the entire process from design to execution. In this scenario, SplitGuard should be implemented at a lower-level, such as the ML libraries the mobile OS supports. However, even in that scenario, the application developer can implement a machine learning pipeline from scratch, without relying on any libraries. This is not an optimal scenario for running SplitGuard. There would have to be strict regulations, as well as gatekeeping by the OS provider (e.g. mandating that machine learning code must use one of the specified libraries) before SplitGuard could effectively be implemented for such clients.

\section{Limitations}

As we have explained in Section \ref{detection}, SplitGuard can potentially, although unlikely, be detected by the attacker, who can then start sending fake gradients from its legitimate surrogate model and regular gradients from its malicious model. This could again cause a significant difference between the fake and regular gradients, and result in a high SplitGuard score. However, a potential weakness of this approach by the attacker is that now the fake gradients result from two different models with different objectives. Suppose the attacker detects SplitGuard at the 200th batch, and starts using its legitimate model. Then the fake gradients within the first 200 batches will be computed using a malicious model, and those after the 200th batch will be computed using the legitimate model. Clients can potentially detect this switch in models, and gain the upper hand. This is another point for which future improvement might be possible.

A final limitation of SplitGuard is that the space of training-hijacking attacks as of this writing is very limited, with only two available related papers \cite{pasquini_unleashing_2021, gawron2022feature}. As more effort is put into this area, it might be possible to develop more sophisticated training-hijacking attacks that cannot be detected by SplitGuard.

\section{Conclusion}

In this paper, we presented SplitGuard, a method for SplitNN clients to detect if they are being targeted by a training-hijacking attack \cite{pasquini_unleashing_2021}. We described the theoretical foundations underlying SplitGuard, experimentally evaluated its effectiveness, and discussed issues related to its use. We conclude that when used appropriately (potentially combined with other tools such as differential privacy \cite{gawron2022feature}), and in a secure setting without label-sharing, a client running SplitGuard can successfully detect training-hijacking attacks and leave the attacker empty-handed. 



\section*{Acknowledgements}
We acknowledge support from T\"{U}B\.{I}TAK, the Scientific and Technological Research Council of Turkey, under project number 119E088. 

\balance
\bibliographystyle{ACM-Reference-Format}
\bibliography{references}

\end{document}